\documentclass[aps,prl,twocolumn,amsmath,amssymb,superscriptaddress,longbibliography]{revtex4-1}

\usepackage{mathrsfs}
\usepackage{graphicx}
\usepackage{amsmath}
\usepackage{amssymb}
\usepackage{amsfonts}
\usepackage{color}
\usepackage{CJK}

\usepackage[unicode=true,colorlinks=true,citecolor=blue,urlcolor=blue]{hyperref}

\def\hatd#1{\hat{#1}^\dagger}
\def\bra#1{\left\langle{#1}\right|}
\def\ket#1{\left|{#1}\right\rangle}
\def\braket#1#2{\left\langle{{#1}}\mathrel{\left|{\vphantom{{#1}{#2}}}\right.\kern-\nulldelimiterspace}{{#2}}\right\rangle}

\begin{document}

\title{Mobility edge of Stark many-body localization}

\author{Li Zhang}

\affiliation{Guangdong Provincial Key Laboratory of Quantum Metrology and Sensing $\&$ School of Physics and Astronomy, Sun Yat-Sen University (Zhuhai Campus), Zhuhai 519082, China}
\affiliation{State Key Laboratory of Optoelectronic Materials and Technologies, Sun Yat-Sen University (Guangzhou Campus), Guangzhou 510275, China}

\author{Yongguan Ke}
\altaffiliation{Email: keyg@mail2.sysu.edu.cn}
\affiliation{Guangdong Provincial Key Laboratory of Quantum Metrology and Sensing $\&$ School of Physics and Astronomy, Sun Yat-Sen University (Zhuhai Campus), Zhuhai 519082, China}
\affiliation{Nonlinear Physics Centre, Research School of Physics, Australian National University, Canberra ACT 2601, Australia}

\author{Wenjie Liu}

\affiliation{Guangdong Provincial Key Laboratory of Quantum Metrology and Sensing $\&$ School of Physics and Astronomy, Sun Yat-Sen University (Zhuhai Campus), Zhuhai 519082, China}
\affiliation{State Key Laboratory of Optoelectronic Materials and Technologies, Sun Yat-Sen University (Guangzhou Campus), Guangzhou 510275, China}

\author{Chaohong Lee}
\altaffiliation{Email: lichaoh2@mail.sysu.edu.cn}

\affiliation{Guangdong Provincial Key Laboratory of Quantum Metrology and Sensing $\&$ School of Physics and Astronomy, Sun Yat-Sen University (Zhuhai Campus), Zhuhai 519082, China}
\affiliation{State Key Laboratory of Optoelectronic Materials and Technologies, Sun Yat-Sen University (Guangzhou Campus), Guangzhou 510275, China}

\begin{abstract}
We investigate many-body localization of interacting spinless fermions in a one-dimensional disordered and tilted lattice.
The fermions undergo \textit{energy-dependent transitions} from ergodic to Stark many-body localization driven by the tilted potential, which are manifested by the appearance of mobility edges between delocalized states and Stark many-body localized states even when the disorder is weak.
We can concretely diagnose these transitions rather than crossovers by finite-size scaling of energy-level statistics.
Moreover, in the Stark many-body localization, the entanglement entropy obeys the area law scaling, in analogy to that in the conventional many-body localization.
\end{abstract}

\date{\today}
\maketitle

\emph{Introduction.} Mobility edge, a critical value separating delocalized and localized states in energy spectrum, has been extensively studied since the seminal work of Anderson localization~\cite{Anderson1958,Mott1975}.
Marking a true quantum phase transition, single-particle mobility edge can exist in three-dimensional random disorder and lower-dimensional quasiperiodic potential~\cite{Abrahams1979,Lee1985,Kondov2011,Semeghini2015,Soukoulis1982,Boers2007,Biddle2011,Ganeshan2015,Li2017,Luschen2018}.
Theoretically, the mobility edge can be determined by the extension or the inverse participation ratio of the wave function of the eigenstates~\cite{Boers2007,Biddle2011,Ganeshan2015,Li2017}.
In experiments, the mobility edge has been measured for noninteracting ultracold atoms in a three-dimensional disordered and a one-dimensional quasiperiodic potential~\cite{Kondov2011,Semeghini2015,Luschen2018}.

Localization of particles in disorder persists even in the presence of interactions, now known as many-body localization (MBL), which serves as a robust mechanism for ergodic breaking in isolated quantum systems~\cite{Gornyi2005,Basko2006,Nandkishore2015,Abanin2019}.
Similar to Anderson transition, there may also exist mobility edge separating many-body localized and ergodic states~\cite{Luitz2015,Serbyn2015}.
MBL has significantly distinct features from ergodic phases, such as logarithmic growth versus ballistic growth of entanglement from nonentangled initial condition~\cite{Znidaric2008,Bardarson2012,Vosk2013,Serbyn2013b}, area law versus volume law scaling of entanglement entropy (EE)~\cite{Bauer2013,Serbyn2013,Kjall2014,Nandkishore2015,Luitz2015,Abanin2019}, and Poisson distribution versus Wigner-Dyson distribution of energy level spacings~\cite{Oganesyan2007,Pal2010,Cuevas2012,Luitz2015}, which can apply to determine mobility edge.
Important experimental progresses have also been made in exploring the MBL in various platforms involving ultracold atoms, trapped ions, nuclear spins, and superconducting circuits~\cite{Smith2016,Kondov2015,Choi2016,Schreiber2015,Bordia2017,Luschen2017,Roushan2017,Xu2018,Wei2018,Rispoli2019,Lukin2019}.

However, disorder is not the essential ingredient for MBL, e.g., a static field may also provide robust mechanism to induce MBL, which is named as Stark MBL~\cite{Schulz2019,Nieuwenbrug2019}.
In closed systems, the Stark MBL shares analogous nonergodic behaviors with the conventional disorder-induced MBL~\cite{Schulz2019,Nieuwenbrug2019,Taylor2020}, such as the Poisson statistic distribution of energy level spacings and the logarithmic growth of entanglement.
A further study shows that the Stark MBL can be distinguished from the conventional MBL via the entanglement growth when the systems are coupled to external dephasing bath~\cite{Wu2019}.
It is natural to ask whether mobility edge exists for interacting particles in a tilted and disordered lattice where Stark MBL is present.
Recently, Ref.~\cite{Chanda2020} found a broad crossover from the ergodic phase to the localized phase as the tilted potential increases.
However, evidence of \textit{energy-dependent transition} from the ergodic phase to the Stark MBL is still lacking.

\begin{figure}[!tp]
\includegraphics[width=0.9\columnwidth]{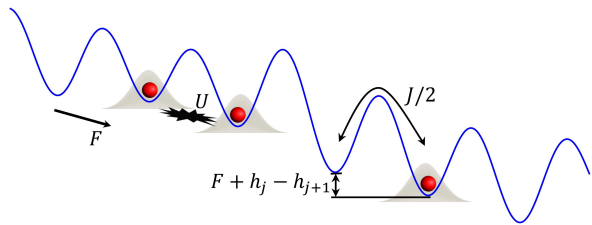}
  \caption{\label{Fig_potential}
  Schematic of the interacting spinless fermions in a disordered lattice subjected to a uniform field with strength $F$.
  The fermions interact through nearest-neighbor interaction $U$ and tunnel between nearest sites with strength $J/2$.}
\end{figure}

In this Rapid Communication, we explore the ergodic-Stark MBL transition of interacting spinless fermions in a one-dimensional optical lattice subjected to uniform field and disorder, as depicted in Fig.~\ref{Fig_potential}.
We calculate the eigenspectrum and eigenstates in an energy-resolved way by using a shift-inverse exact diagonalization method.
We uncover a clear mobility edge through analyzing the statistics of the eigenspectrum and finite-size scaling.
Furthermore, we study the entanglement structure of the eigenstates and find that the EE obeys the area law scaling in the Stark MBL.
Our result lays a concrete cornerstone for detecting mobility edge of Stark MBL in future experiments.
%
%
%

\emph{Disordered Stark ladder of interacting spinless fermions.} We consider an ensemble of interacting spinless fermions in a one dimensional disordered lattice in the presence of a uniform field; see Fig.~\ref{Fig_potential}, which is described by the Hamiltonian
\begin{equation}
\hat H=\sum_{j=1}^{L-1}\left[\frac{J}{2}\left(\hatd c_j\hat c_{j+1}+\hatd c_{j+1}\hat c_j\right)+U\hat n_j\hat n_{j+1}\right]+\sum_{j=1}^LV_j\hat n_j.
\end{equation}
Here, the system size is finite with total number of sites $L$.
$\hatd c_j (\hat c_j)$ creates (annihilates) a fermion at site $j$ and $\hat n_j=\hatd c_j\hat c_j$ is the particle number operator.
$\frac{J}{2}$ and $U$ are the nearest-neighbor tunneling and interaction strength, respectively.  $V_j=h_j-Fj$ is the position-dependent potential, where $F$ is the strength of the uniform field and $h_j\in\left[-W,W\right]$ is uniformly distributed with disorder strength $W$.
The energy unit is set as $J=U=1$.
In the absence of disorder and at single-particle level, the eigenstates are the well-known Wannier-Stark states and the energy spectrum forms an equidistant ladder, namely the Stark ladder~\cite{Wannier1960,Wannier1962,Fukuyama1973}.
The Stark ladder causes exact many-fold degeneracies in the many-body energy spectrum, which can be lifted by the disorder to recover the generic localization behaviors~\cite{Nieuwenbrug2019,Taylor2020}.

We consider half-filling of the lattice with total particle number $N=L/2$ in the open boundary condition.
By exact diagonalization method, we study such system in different sizes by averaging different disordered configurations: $L=12$ (at least 2000 samples), $14$ (1000 samples), $16$ (1000 samples), and $18$ (600 samples).
Using the shift-invert spectral transformation $(\hat H-E\hat I)^{-1}$ with $\hat I$ being the identity matrix, we can reach eigenstates at any energy density $\epsilon=(E-E_{\mathrm{min}})/(E_{\mathrm{max}}-E_{\mathrm{min}})$~\cite{Luitz2015}, where $E_{\mathrm{min}}$ and $E_{\mathrm{max}}$ are the minimum and maximum eigenergies for each disorder realization, respectively.
In the following studies, we consider $\epsilon\in[0.15, 0.85]$, since the densities of states are too low at the high- and low-energy tails of the energy spectrum.
For each parameter point and each disorder realization, we take the closest 50 eigenpairs around each $\epsilon$.
The observables are calculated from the corresponding eigenstates and averaged over the set of eigenpairs and disorder realizations.

\begin{figure}[!bp]
\includegraphics[width=0.8\columnwidth]{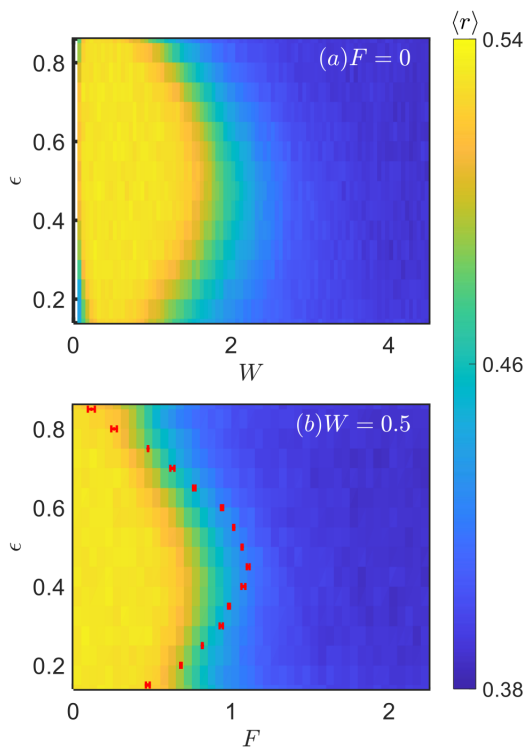}
  \caption{\label{Fig_PD}
  (a) The disorder-averaged ratio of energy level spacings $\langle r\rangle$ as a function of $(\epsilon, W)$ at $F=0$.
  In the weak disorder region, all the eigenstates are ergodic.
  (b)  $\langle r\rangle$ as a function of $(\epsilon, F)$ at $W=0.5$.
  The colors denote the values of $\langle r\rangle$ at $L=16$.
  %
  The numerical data marked by red dots are the critical points extracted from true finite-size scaling of $\langle r\rangle$ at system size $L=12$, $14$, $16$, and $18$.
  The error bars represent the standard deviation from different widths of fitting window used in the scaling process.
  }
\end{figure}

\begin{figure*}[!htbp]
\includegraphics[width=2\columnwidth]{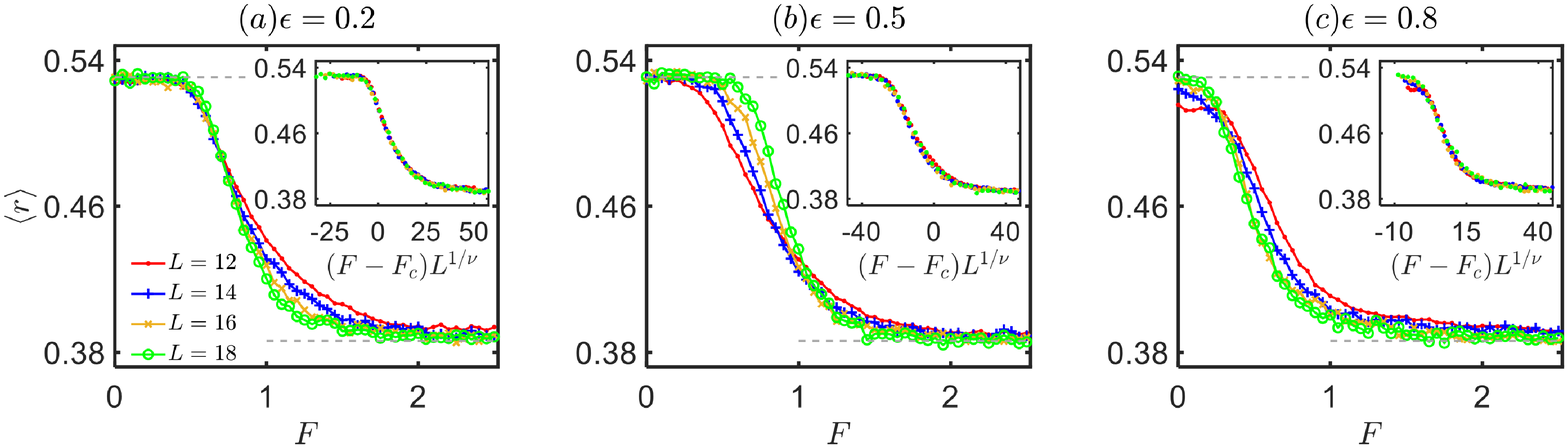}
  \caption{\label{Fig_LevelSta}
  The disorder-averaged ratio of the consecutive level spacings $\langle r\rangle$ as a function of the field strength at different energy densities (a) $\epsilon=0.2$, (b) $\epsilon=0.5$, and (c) $\epsilon=0.8$.
  The data at system sizes $L=12$, 14, 16, and 18 are denoted by red lines marked by dots, blue lines marked by pluses, yellow lines marked by crosses, and green lines marked by circles, respectively.
  The two dashed gray lines correspond to $\langle r\rangle_{\mathrm{GOE}}$ and $\langle r\rangle_{\mathrm{P}}$ respectively.
  Inset: collapses of the data used to extract the critical values: (a) $F_c\approx0.683$, $\nu\approx 0.75$, (b) $F_c\approx1.066$, $\nu\approx 0.7$, and (c) $F_c\approx0.256$, $\nu\approx0.82$.
  The width of the fitting window corresponds to $w=1$.
  }
\end{figure*}

\emph{Many-body mobility edge induced by uniform field.} A well-known criterion to distinguish the MBL from the ergodic phase is the statistic of the many-body eigenspectrum.
The ratio of adjacent energy gap is defined as
\begin{equation}
r_n=\min{\left(\delta_{n+1}/\delta_{n},\delta_n/\delta_{n+1}\right)},
\end{equation}
with $\delta_n=E_{n}-E_{n-1}$ being the energy gap between eigenergy $E_{n-1}$ and $E_n$.
In the many-body localized system such ratio obeys the Poisson distribution, $P_{\mathrm{P}}(r)=2/(1+r)^2$, while in the ergodic system it obeys the Wigner-Dyson distribution of Gaussian orthogonal ensembles, $P_{\mathrm{GOE}}(r)=8\left(r+r^2\right)/27\left(1+r+r^2\right)^{2.5}$~\cite{Atas2013}.
It is convenient to use $\langle r\rangle$, the average of adjacent gap ratio over different disorder realizations, to distinguish the two phases.
$\langle r\rangle$ changes from $\langle r\rangle_{\mathrm{P}}\approx0.386$ in the localized phase to $\langle r\rangle_{\mathrm{GOE}}\approx 0.531$ in the ergodic phase.
In the absence of the field, all the eigenstates are ergodic when $W\lesssim1.8$~\cite{Luitz2015}.
In Fig.~\ref{Fig_PD}(a), we show how the disorder-averaged ratio $\langle r\rangle$ changes with energy density and disorder strength at $F=0$ and system size $L=16$.
We numerically checked the values of $\langle r\rangle$, and find that $\langle r\rangle\approx \langle r\rangle_{\mathrm{GOE}}$ for all energy densities when $0.1\lesssim W\lesssim1$.
Without loss of generality, in the following study we fix $W=0.5$, at which all the eigenstates at $F=0$ are ergodic, and explore whether the field will induce transition from the ergodic phase to the localized phase.

In Fig.~\ref{Fig_PD}(b), we show the distribution of the disorder-averaged gap ratio $\langle r\rangle$ in the plane of $\left(F, \epsilon\right)$ at $W=0.5$ and $L=16$.
It is obvious that for all the energy densities, $\langle r\rangle$ crossovers from $\langle r\rangle_{\mathrm{GOE}}$ to $\langle r\rangle_{\mathrm{P}}$ as the strength of the field increases, which indicates that there may be a transition from the ergodic phase to the Stark MBL.
Moreover, there also exists crossover from the ergodic phase to the localized phase as energy changes, which indicates the possible existence of a mobility edge.

To ascertain the energy-dependent ergodic-Stark MBL transition and the critical points, we perform finite-size scalings of $\langle r\rangle$ for different energy densities, and the system sizes are chosen as $L=12$, 14, 16, and 18.
In Fig.~\ref{Fig_LevelSta}, we plot $\langle r\rangle$ as a function of $F$ at energy densities $\epsilon=0.2$, 0.5, and 0.8.
As the strength of the field increases, the values of $\langle r\rangle$ for different system sizes cross around a critical value $F_c$ between the two limiting values $\langle r\rangle_{\mathrm{GOE}}$ and $\langle r\rangle_{\mathrm{P}}$.
The critical points $F_c$ are extracted via rescaling $F$ by $\left(F-F_c\right)L^{1/\nu}$, so that the data of $\langle r\rangle$ for different system sizes collapse into a single universal function of $\left(F-F_c\right)L^{1/\nu}$, see the insets in Fig.~\ref{Fig_LevelSta}.
The critical values for different energy densities are plotted in Fig.~\ref{Fig_PD}(b) (see the red dots with error bars), which shows a clear mobility edge induced by the uniform field.
We notice that this mobility edge is asymmetric about the center of the energy spectrum: it is slightly shifted towards the ground energy, which resembles that in the disordered many-body systems~\cite{Luitz2015,Serbyn2015}.
Our result shows that the maximum critical point is around $F_c\approx1.11$ at $\epsilon=0.45$, which is smaller than the critical point of $F_c\approx2.2$~\cite{Nieuwenbrug2019} obtained from considering the statistic of the entire eigenspectrum under period boundary condition.
We attribute this deviation to the difference between open and period boundary conditions.
The connectivity of the system is lower under open boundary condition, thus the transition points to the localized phase should be smaller.

We now present the details of extracting the critical values $F_c$ in the collapse process~\cite{Nieuwenbrug2019}.
First, we rescale all the curves at different system sizes by transforming $F$ into $x=\left(F-F_c\right)L^{1/\nu}$ with proposed sets of $\left(F_c, \nu\right)$.
Then, we optimize $\left(F_c,\nu\right)$ through minimizing the difference of the rescaled curves $\mathcal{D}(F_c,\nu)=\frac{1}{2wR}\sum_{i<j}\int_{-wR}^{wR}\left(y_i(x)-y_j(x)\right)^2dx$, where $i$, $j=$12, 14, 16, 18, and $y_i(x)$ represents the spline-interpolated data of the curves.
The range of the integration is given by $[-wR,wR]$, where $R$ is set as $\max{[(F-F_c)12^{1/\nu}]}-\min{[(F-F_c)12^{1/\nu}]}$ and $w$ defines the width of the fitting window used for the collapse process, for which we test from 0.1 to 1.
Finally, we extract $F_c$ and $\nu$ by averaging over all the chosen $w$ and give the error bars by the standard deviation.
We notice that the error bar is relatively large at the top of the energy spectrum.
It can be ascribed by the severe finite size effect, as can be seen in Fig.~\ref{Fig_LevelSta}(c), where the value of $\langle r\rangle$ around $F=0$ at $L=12$ deviates from $\langle r\rangle_{\mathrm{GOE}}$ severely.

\begin{figure*}[!htbp]
	\includegraphics[width=2\columnwidth]{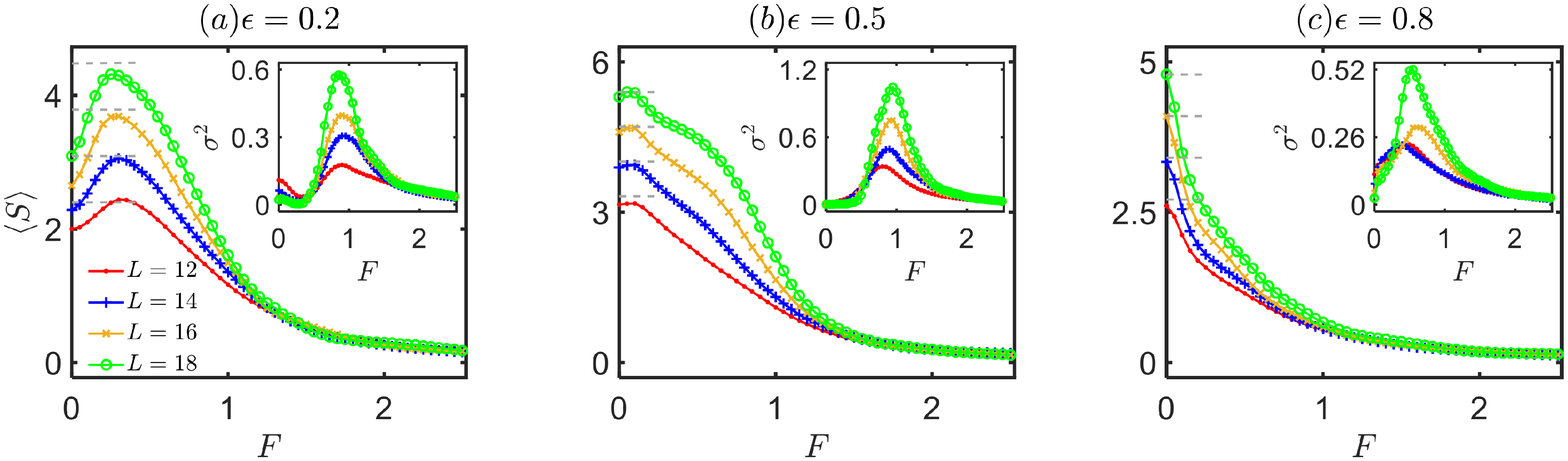}
	\caption{\label{Fig_Entropy_var}
		The disorder-averaged von Neuman EE $\langle S\rangle$ as a function of the field strength $F$ at different energy densities (a) $\epsilon=0.2$, (b) $\epsilon=0.5$, and (c) $\epsilon=0.8$.
		The data at system sizes $L=12$, 14, 16, and 18 are denoted by red lines marked by dots, blue lines marked by pluses, yellow lines marked by crosses, and green lines marked by circles, respectively.
		The dashed gray lines correspond to the maximal values of EE: (a) $\frac{L}{2}\ln2-1.76$, (b) $\frac{L}{2}\ln2-0.838$, and (c) $\frac{L}{2}\ln2-1.445$.
		Insets: the disorder-averaged variance $\sigma^2$ of the von Neuman EE as a function of the field strength.
		The peaks become sharper as the system size increases and the peak locations indicate the ergodic-Stark MBL transition points.
	}
\end{figure*}

At last, we study the entanglement properties of the many-body eigenstates in the ergodic phase and the Stark MBL.
The bipartite entanglement between a subsystem A and the rest of the system B can be quantified by the von Neuman EE
\begin{equation}
S=-\mathrm{Tr}\left(\hat \rho_{\mathrm{A}}\ln\hat \rho_{\mathrm{A}}\right),
\end{equation}
where $\hat \rho_\mathrm{A}$ is the reduced density matrix of the subsystem A.
We consider the half-chain EE, i.e. the subsystem A is chosen to be a block of $\frac{L}{2}$ contiguous sites.
In our calculation, the eigenstates are expressed in the Fock bases as $\ket{\psi}=\sum_{n_1\cdots n_L}C_{n_1\cdots n_L}\ket{n_1\cdots n_L}$, where $n_i=0$, 1 is the number of particles on site $i$ and $C_{n_1\cdots n_L}$ is the complex amplitude.
The reduced density matrix of the subsystem A then reads
\begin{equation}
\hat \rho_{\mathrm{A}}=\sum_{\vec{n}_{\mathrm{L}},\vec{n}'_{\mathrm{L}}}\ket{\vec{n}_{\mathrm{L}}}\bra{\vec{n}'_{\mathrm{L}}}\sum_{\vec{n}_{\mathrm{R}}}C_{[\vec{n}_{\mathrm{L}},\vec{n}_{\mathrm{R}}]}C_{[\vec{n}'_{\mathrm{L}},\vec{n}_{\mathrm{R}}]}^*,
\end{equation}
where $\vec{n}_{\mathrm{L}}=\{n_1\cdots n_{L/2}\}$ and $\vec{n}_{\mathrm{R}}=\{n_{L/2+1}\cdots n_{L}\}$ are shorted for the sequence of occupation numbers of the left and right part of the chain.
According to the above reduced density matrix, we can calculate the von Neuman EE for different parameters and disorder realizations.
In the ergodic phase, the eigenstates are thermal, thus the von Neuman EE equals to the thermodynamic entropy, which is extensive.
This leads to the volume law scaling of the von Neuman EE, that is, $S$ is proportional to the volume of the subsystem A~\cite{Nandkishore2015,Abanin2019}.
In the disorder induced MBL, the eigenstats can be written as product states by quasilocal unitary transformations, which implies that the von Neuman EE is proportional to the area of the surface between the two subsystems, that is, it obeys the area law scaling~\cite{Bauer2013,Serbyn2013,Kjall2014,Nandkishore2015,Luitz2015,Abanin2019}.

We now test the volume law and area law scalings of the EE in the ergodic phase and the Stark MBL respectively.
The disorder-averaged $\langle S\rangle$ is shown in Fig.~\ref{Fig_Entropy_var}, as a function of $F$ at energy densities $\epsilon=0.2$, 0.5, and 0.8 and different system sizes.
In the ergodic phase at weak field strength, $\langle S\rangle$ is size-dependent and behaves as $Lf_\epsilon(F)+c_\epsilon(F)$, signifying the volume law scaling.
The maximum EE at each energy density approaches $\frac{L}{2}\ln2+c_\epsilon$, as shown by the dashed gray lines in Fig.~\ref{Fig_Entropy_var}.
On the other hand, in the localized phase at strong field, the averaged EE is much lower than that in the ergodic phase and collapses into a single curve for all system sizes, meaning that the EE obeys the area law scaling in the Stark MBL.

In the insets of Fig.~\ref{Fig_Entropy_var}, we plot the variance of the EE $\sigma^2=\langle S^2\rangle-\langle S\rangle^2$ as a function of the field strength at energy densities $\epsilon=0.2$, 0.5, and 0.8.
It is well known that for the ergodic-MBL transition, the variances of the EE in finite systems show peaks near the transition points, originating from the coexistence of the delocalized and localized regime~\cite{Kjall2014}, and can be used to locate the transition points~\cite{Kjall2014,Luitz2015}.
We find that this scenario is also valid in the ergodic-Stark MBL transition, as can be seen in the insets of Fig.~\ref{Fig_Entropy_var}, where the peaks of the variance become sharper and the corresponding locations tend to the transition points as the system size increases.

\emph{Conclusion.} We have revealed the existence of a mobility edge between ergodic and Stark MBL states of interacting spinless fermions in a disordered Stark ladder.
We find that the spectral properties of the Stark MBL are in common with that of the conventional disorder induced MBL.
Specifically, the statistic of the adjacent gap ratio of the eigenspectrum is of Poisson type and the EE obeys the area-law scaling in the Stark MBL.
Through a finite-size scaling of the adjacent gap ratio, we give the phase diagram of the mobility edge in the eigenspectrum.

Our findings of the mobility edge can be tested experimentally with ultracold atoms in a tilted lattice by employing the methods in Ref.~\cite{Luschen2018}.
We note that in our work, the interaction strength is fixed at a relatively small value.
In a one-dimensional disordered many-body system, it has been shown that increasing the interaction strength would cause a reentrance into the localized phase at weak disorder strength~\cite{Bar2015}.
It is interesting to study how the interplay between the interaction and the uniform field affects the localization, explicitly the resonant tunneling effect on the Stark MBL, where tunneling is enhanced when the interaction and the bias field match.
We expect it will induce new phenomena that are absent in the disordered system, and provide a way to distinguish MBL and Stark MBL.

\emph {Note added}. Recently, we became aware of the related work of studying Stark MBL of bosons and the coexistence of localized and ergodic states in harmonic trap~\cite{Yao2020}.

\acknowledgments
This work is supported by the Key-Area Research and Development Program of GuangDong Province under Grants No. 2019B030330001, the National Natural Science Foundation of China (NNSFC) under Grants No. 11874434 and No. 11574405, and the Science and Technology Program of Guangzhou (China) under Grants No. 201904020024. Y.K. is partially supported by the Office of China Postdoctoral Council (Grant No. 20180052), the National Natural Science Foundation of China (Grant No. 11904419), and the Australian Research Council (DP200101168).


\end{document}